# VisAhoi: Towards a Library to Generate and Integrate Visualization Onboarding Using High-level Visualization Grammars

Christina Stoiber[a,*], Daniela Moitzi[b], Holger Stitz[b], Florian Grassinger[a], Anto Silviya Geo Prakash[b], Dominic Girardi[b], Marc Streit[c] and Wolfgang Aigner[a]

[a]*St. Pölten University of Applied Sciences, St. Pölten, Austria*
[b]*datavisyn GmbH, Linz, Austria*
[c]*Johannes Kepler University, Linz, Austria*

ABSTRACT

Visualization onboarding supports users in reading, interpreting, and extracting information from visual data representations. General-purpose onboarding tools and libraries are applicable for explaining a wide range of graphical user interfaces but cannot handle specific visualization requirements. This paper describes a first step towards developing an onboarding library called VisAhoi, which is easy to *integrate, extend, semi-automate, reuse, and customize*. VisAhoi supports the creation of onboarding elements for different visualization types and datasets. We demonstrate how to extract and describe onboarding instructions using three well-known high-level descriptive visualization grammars—Vega-Lite, Plotly.js, and ECharts. We show the applicability of our library by performing two usage scenarios that describe the integration of VisAhoi into a VA tool for the analysis of high-throughput screening (HTS) data and, second, into a Flourish template to provide an authoring tool for data journalists for a treemap visualization. We provide a supplementary website (https://datavisyn.github.io/visAhoi/) that demonstrates the applicability of VisAhoi to various visualizations, including a bar chart, a horizon graph, a change matrix/heatmap, a scatterplot, and a treemap visualization.

## 1. Introduction

Data visualization has become an indispensable tool not only in the context of science and business but also in everyday life, such as data stories in newspapers, in books, on the internet, or personal data (e.g., sleep tracking, nutrition, sports, etc.). Most recently, the complexity and social relevance of the COVID-19 pandemic has put data visualization at the center of worldwide attention [64]. Since the outbreak, data visualization researchers and experts have been providing various data visualizations for public education. This way, the general public got in touch with diverse data visualizations presenting medical data such as reproduction numbers, COVID-19 cases, hospitalization, etc. Visualization onboarding methods [66, 70] aim to support end users in comprehending data visualizations and taking full advantage of the tools at hand. However, more research on onboarding concepts for visualization techniques and visual analytics tools is needed. In current literature, different onboarding approaches have been investigated in the context of various visualization techniques, including parallel coordinates plots [37, 56], hyperbox [56], spiral charts [56], treemaps [56, 75], network graphs [75], and storylines [75] were investigated using different onboarding approaches.

Concerning available products for user onboarding, solutions mostly focus on the overall user interface rather than the visual representations in particular (e.g., Appcues [4] or Intercom [33]). Most web-based user onboarding libraries [31, 5, 34, 78, 76, 48, 27, 77] cannot be readily applied for visualization onboarding as they are limited to guided tours explaining elements in the user interface, which block the whole interface. Especially when exploring visualizations, it is crucial to allow users to interact with the visualization and onboarding simultaneously, not to block

*Corresponding author

✉ firstname.lastname@fhstp.ac.at (C. Stoiber); fistname.lastname@datavisyn.io (D. Moitzi); fistname.lastname@datavisyn.io (H. Stitz); firstname.lastname@fhstp.ac.at (F. Grassinger); fistname.lastname@datavisyn.io (A.S.G. Prakash); fistname.lastname@datavisyn.io (D. Girardi); marc.streit@jku.at (M. Streit); firstname.lastname@fhtp.ac.at (W. Aigner)

ORCID(s): 0000-0002-1764-1467 (C. Stoiber); 0009-0006-6686-8429 (D. Moitzi); 0000-0002-4742-2636 (H. Stitz); 0000-0003-4409-788X (F. Grassinger); 0000-0001-7511-2910 (A.S.G. Prakash); 0009-0008-1085-3433 (D. Girardi); 0000-0001-9186-2092 (M. Streit); 0000-0001-5762-1869 (W. Aigner)

the whole interface when seeking help. While existing learning environments and web platforms such as "The graphic continuum" [72] or the "Data Viz Catalogue" [55] focus on the explanation of visualization techniques, the presented textual descriptions and illustrations are mainly manually developed by the visualization designer of the respective platform. These onboarding solutions are generally tailored to specific visualization techniques and datasets or are restricted in reusability or integration capabilities into existing visualization tools.

In this work, we aim to create a versatile and reusable solution for visualization designers and developers to generate visualization onboarding instructions semi-automatically with textual descriptions (onboarding messages) and in-place annotation anchors (see Figure 1 and 7) that allow the users of a visual analytics (VA) tool to interact with the visualization while interacting with the onboarding instructions. This allows the user to self-explore the onboarding instructions and receive context-sensitive help at any point of their exploration with the VA tool. Furthermore, our onboarding solution is designed to be reusable and can be integrated into custom visualizations (see Figure 1). The goal is not to create static visualization tutorials but to enable the developers to integrate the onboarding seamlessly into the productively used visualizations. The end users can switch on the onboarding as they need it directly in their currently used visualization with the actual loaded data. To achieve this, we leverage high-level declarative visualization grammars, which aim to simplify programming tasks by specifying "what" visualization results should be rather than "how" they should be computed [58]. They can simplify visualization specifications while supporting a high degree of expressiveness and customization. We use the power of high-level declarative visualization grammars to extract and describe visualization onboarding instructions.

We propose a JavaScript library called *VisAhoi* and show the integration exemplary for the three visualization libraries with high-level declarative grammars—Plotly.js [50], Vega-Lite [57], and Apache ECharts [41]—to explore how we can semi-automatically generate onboarding instructions based on the currently shown dataset. Following the contribution types of visualization research as defined in [45], our work falls into the category of a systems paper. Our contributions include:

- A description of our **onboarding concept** (see Sec. 2.4), which we iteratively developed and revised based on previous studies [70, 65] and feedback of domain experts.
- The description of the **design and architecture of the VisAhoi** JavaScript library in Sec. 5. We discuss extracting information regarding the visual encoding of the high-level declarative description, the JavaScript runtime object, and the DOM node.
- A demonstration of the feasibility of our approach by performing two **usage scenarios** [59] (see Sec. 6). The first usage scenario presents the integration of VisAhoi into a VA tool for analyzing high-throughput screening (HTS) data for biomedical R&D, and second, into a Flourish template to provide an authoring tool for data-journalistic use cases. We also show the feasibility of our approach for a bar chart, a change matrix [46], a horizon graph [30], scatterplot, and treemap [63], which can be accessed here: `https://datavisyn.github.io/visAhoi/demos/`. The design of the semi-automatically generated onboarding instructions is shown in a horizon graph, illustrated in Figure 7 in this paper.
- A discussion on **limitations** of the current approach in Sec. 6.2.

## 2. Related Work

### 2.1. Information Extraction from Visualizations

Visualization onboarding requires information about the visualization to create, for instance, tailored messages or set the position of the annotations. A body of previous work describes approaches to extract information from visualizations as bitmap images to reverse engineer them [51], create re-usable visualization templates that can be used with different datasets [29], or create textual summaries for visually impaired people [47]. Most techniques employ machine learning algorithms to extract structural information (e.g., axes, marks), visual encoding, and labels from the visualization.

In contrast, Harper and Agrawala [28] present an approach to deconstructing D3 charts to extract their underlying structure, i.e., data, marks, and the mapping between them. Similar to our approach, they focus on SVG-based visualizations but extract the information directly from the DOM structure and data attached to the DOM nodes. This method leads to limitations when, for instance, deconstructing non-linear functional mappings, e.g., log scale in a scatterplot. In *VisAhoi*, we employ the declarative description, the JavaScript runtime object, and the DOM structure. As a result, we have access to internal information and functions of the visualization library, which can be used for tailored onboarding messages.

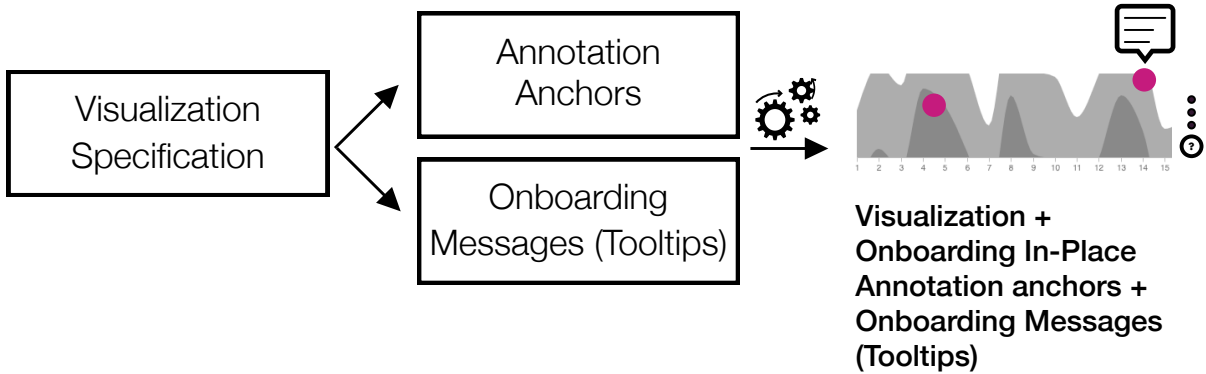

Figure 1: **Workflow for visualization developers:** Based on a visualization specification (e.g., Plotly.js [50], Vega-Lite [57], etc.) and the definition of a visualization type (e.g., bar chart, change matrix, etc.), the VisAhoi Library semi-automatically generates in-place annotation anchors and onboarding messages in the form of tooltips.

### 2.2. Visualization Grammar

Most recently, McNutt [44] surveyed JSON-style declarative domain-specific languages over the last 20 years. One of the first examples of a grammar-based approach to visualization was Wilkinson's *The Grammar of Graphics* [81], an abstraction that makes thinking, reasoning, and communicating about graphics easier. Wickham's popular *ggplot2* [79] and *ggvis* [80] packages implement variants of Wilkinson's model in the R statistical language. Furthermore, Shih et al. [62] developed a declarative grammar to specify advanced volume visualizations effectively. ggplot2 and ggvis provide specific, pre-defined types of visualizations. In contrast, languages such as *Protovis* [14] or *Vega* [58] provide basic building blocks for a wide variety of visualization designs: data loading and transformation, scales, map projections, axes, legends, and graphical marks such as rectangles, lines, plotting symbols, etc. Interaction techniques can be specified using reactive signals that dynamically modify a visualization in response to input event streams. *Vega-Lite* [57] expands the basic interaction features of Vega and provides a high-level specification of interactive data visualizations. It enables specifying visualizations in terms of marks, layout, and data. *Encodeable* [82] is a new configurable grammar independent from rendering where a component author can declare a grammar for the encoding channel of their component, which looks like a subset of the Vega-Lite grammar. Li et al. [41] introduced *ECharts* as a web-based framework for rapidly constructing cross-platform visualizations. The ECharts library employs an all-in-one JSON format option to declare the components, styles, data, and interactions, resulting in a logicless and stateless mode. *Plotly.js* is also a high-level, declarative charting library that is highly used and covers 40 visualization types, including 3D charts, statistical graphs, and SVG maps [50].

To validate the applicability of our approach, we selected the visualization libraries Vega-Lite, Apache ECharts, and Plotly.js to prototypically demonstrate the extraction of information from declarative visualization specifications and generate onboarding instructions in an automated manner. We decided to use those three visualization grammars due to the following characteristics: (1) all libraries are **high-level declarative visualization grammars**, (2) we explicitly target visualization libraries with **SVG** rendering, as it is important for employing our onboarding annotations in the DOM structure (see Section 5.1 for more detail), and (3) all libraries are **open-source**.

### 2.3. Visualization Onboarding

Visualization onboarding methods [70] aim to support end users in comprehending data visualizations and taking full advantage of the tools at hand. So far, there has been little discussion about onboarding concepts for visualization techniques and VA tools [22, 69]. The educational community started by studying how students interpret and generate data visualizations [6, 2, 9]. More recently, Firat et al. [21] developed an interactive pedagogical treemap application for training. Additionally, Peng et al. [49] present results of a study to evaluate six parallel coordinate literacy modules based on Bloom's taxonomy [10] using videos, tests, and tasks.

In the current literature, Tanahashi et al. [75] investigated *top-down* and *bottom-up teaching methods* and *active* or *passive learning types*. The bottom-up teaching method ("textbook approach") [83] focuses on small, detailed pieces of information that students then combine to get a better understanding. Conversely, a top-down teaching method is given when a broad overview first helps to understand the abstract, high-level parts of an idea/topic, providing context for understanding its components in detail [75]. Furthermore, a distinction can be made between active and passive learning types. Passive learning means that students only receive the information without participatory dialog. In contrast, active learning describes an active participation [75].

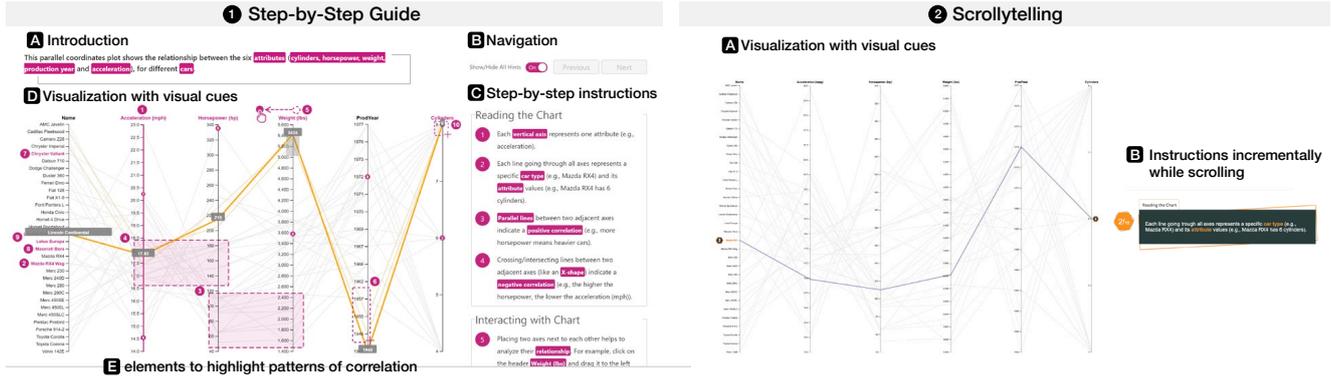

**Figure 2:** ❶ **Step-by-step guide for a Parallel Coordinates Plot.** The step-by-step guide is based on textual descriptions and in-place annotations anchors. It consists of four parts: a brief textual introduction providing contextual information about the visualization Ⓐ, navigational elements Ⓑ to go through the step-by-step instructions Ⓒ, and the visualization itself Ⓓ. We divide the textual descriptions into *Reading the Chart*, for explaining the visual encoding, *Interacting with the Chart*, for explaining the interaction concept, and *Using the Chart*, for providing insights. ❷ **Scrollytelling tutorial.** Users can incrementally scroll through the instructions Ⓑ while the visualization Ⓐ changes to the text appropriately.

In their comparative study, Known and Lee [37] explored the effectiveness of *active learning* strategies. Three tutorial types—static, video-based, and interactive—were used to support the learning of parallel coordinates plot visualizations. They observed that participants who used interactive and video tutorials outperformed those who used static or no tutorials. Their analysis indicates that top-down exercises were more effective than bottom-up and active learning types, with top-down tasks being the most effective.

Ruchikachron et al. [56] found out that the *learning by analogy* concept is helpful as participants in their study could entirely or at least significantly understand the unfamiliar visualization methods better after they observed the transitions from a familiar counterpart. They assessed four combinations: scatterplot matrix against hyperbox, linear chart against the spiral chart, hierarchical pie chart against treemap, and data table against parallel coordinates plots. Hyperbox is a scatterplot matrix that allows non-orthogonal axes [3]. The authors also describe another advantage of learning-by-analogy over other forms of demonstrations, such as textual or oral descriptions: the power of visuals, as they bridge any language barriers.

Stoiber et al. [70] developed four onboarding concepts (a step-by-step guide, scrollytelling, a video tutorial, and an in-situ scrollytelling), see Figure 2. They conducted two quantitative comparative user studies with MTurk workers and a qualitative comparative user study with students. The main aim of these studies was to investigate the effect of onboarding on user performance and evaluate the subjective user experience. They proposed guidelines for the design of visualization onboarding methods, in particular, (1) onboarding systems should be integrated into the visualization tool appreciated; (2) use of an easy-to-understand data set and concrete examples on how to read the charts is vital, they support and increase comprehension; (3) to-the-point descriptions make it easier to absorb information (step-by-step); (4) Some users tend to ignore onboarding systems, even if they struggle. These users have to be motivated to use onboarding.

Additionally, the authors explored abstract and concrete onboarding instructions for a treemap visualization and assessed them in a quantitative comparative user study with students [65]. The results reflect the discussion of abstract or concrete materials for teaching [19, 35]. The authors distinguish between **concrete** and **abstract** onboarding messages, while the former tries to include concrete facts from the visualizations, e.g., values or insights — *"The size of each rectangle represents the spending in US-Dollars (e.g., the Housing rectangle, representing $213 billion, is approx. twice as large as the Clean drinking rectangle, representing $111 billion.)"* The latter tries to provide a more generic description of the treemap, which doesn't include any concrete values from the visualization, e.g., *"The size of each rectangle represents a quantitative value associated with each element in the hierarchy."* The study with

VisAhoi: A Library to Generate Visualization Onboarding

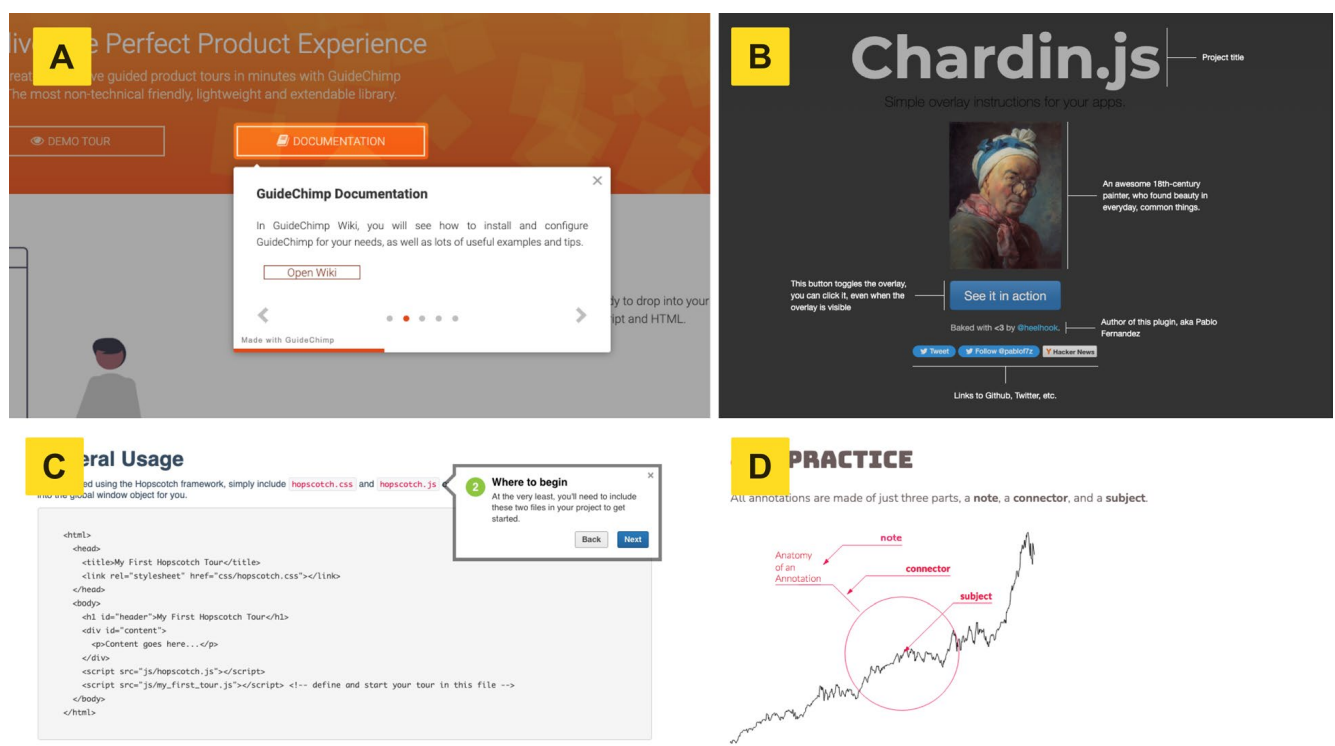

**Figure 3: Examples of user and visualization onboarding JavaScript libraries** (A) GuideChimp [27] providing guided tours with step-wise instructions. (B) chardin.js [16] uses static labels to onboard different user interface elements. (C) Hopscotch [31] uses guided tours with buttons to navigate the steps. And (D) the d3-annotation library by [42] integrates labels that can be positioned in the visualization.

40 participants revealed the following results: (1) Concrete onboarding messages, which refer to the underlying data, are more helpful than abstract messages, whereas the shorter length of the abstract messages is perceived to be more appropriate. (2) Both concrete and abstract onboarding messages can lead to precious insights.

Along with our derived lessons learned, we further improved our onboarding design by creating a high-fidelity clickable prototype using Sketch[1]. Thus, we present the resulting **visualization onboarding design concept** in our paper [67]. We integrated our onboarding concept exemplarily in a VA tool to analyze high-throughput screening (HTS) data. The HTS analysis application holds data from HTS experiments where the binding kinetics of roughly 4 Mio. chemical compounds against potential drug targets are measured. The measurement for each compound is a time series curve that describes the binding behavior of the compound against the target over time. The 4 Mio time series measurements are projected into a two-dimensional latent space with a multidimensional scaling approach. The 2D projection is visualized with a scatterplot. Each dot represents a single compound, and the topology in the 2D space reflects the topology of the time series curves in the high-dimension space. Domain experts search for clusters in these scatterplots with compounds showing promising binding behavior. Two scatterplots visualize the data side-by-side. Therefore, we target our onboarding concept to this VA tool's scatterplot visualizations and users. We evaluated our onboarding design by conducting a cognitive walkthrough and interviews with thinking aloud. We also collected data on domain experts' visualization literacy. The results of the cognitive walkthrough showed that domain experts positively commented on the onboarding design and proposed adjusting smaller aspects. The interviews revealed that domain experts are well-trained in interpreting basic visualizations (e.g., scatterplot, bar chart, line chart). However, they need support correctly interpreting the data visualized in the scatterplot, as they were new to them. Another important insight was fitting the onboarding messages into the domain's language.

---

[1] https://www.sketch.com, Accessed: 2023-08-29

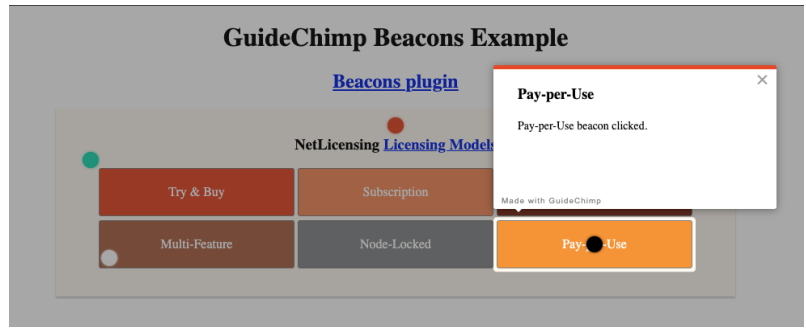

**Figure 4: Self-exploratory animated beacons** provided by the library GuideChimp [27].

## 2.4. User Onboarding Tools and Libraries

Several commercial software tools are available for integrating onboarding elements into user interfaces. For instance, Appcues [4], a tutorial-based onboarding service, can provide contextual help. Besides web applications, some frameworks and JavaScript libraries can be used to onboard users. Most libraries provide guided tours to introduce users to an application. Hopscotch [31] is a powerful framework for building a simple welcome tour (illustrated in Figure 3 (C)) using an exchangeable JSON format system. Other libraries for guided tours are aSimpleTour [5], intro.js [34], tourguide [76], webTour.js [78], tooltip sequence [60], shepherd.js [61], orient [48], GuideChimp [27], tutoBox [77], and driver.js [20], which allow the integration of animated guided tours. All those libraries integrate buttons and/or step-by-step widget interaction elements with a strict pre-defined sequence of steps for the users. Therefore, supporting the self-exploratory approach we propose with our VisAhoi library is impossible, as there is a strict pre-defined sequence of steps. Other available libraries, like chardin.js [16], provide visually appealing labels but no interactive elements as shown in Figure 3 (B). The presented libraries are feasible for onboarding different web applications but are limited to guided tours and static labels. Therefore, it is unusable for our proposed visualization onboarding concepts with in-place annotations and onboarding instructions.

Additionally, most onboarding libraries do not allow users to interact with the underlying interface while reading the onboarding instructions. Especially when exploring visualizations, it is crucial to allow users to interact with the visualization and onboarding simultaneously without blocking the whole interface. VisAhoi allows users to open the onboarding instructions at any point (self-exploratory). The presented JavaScript libraries [31, 5, 34, 78, 76, 48, 27, 77] restrict users to step-wise navigate (guide) through the onboarding instructions only in a pre-defined, fixed sequence. With GuideChimp [27], developers can extend the basic library with a plugin to also provide animated beacons (see Figure 4) in the interface to open the onboarding instructions in a self-exploratory way. Moreover, after closing the onboarding, users have to start from scratch. So, the onboarding libraries don't provide context-sensitive help, as users must navigate through all the instructions to find the most related instruction to their problem.

To address the specific needs of visualization tools, there are also some visualization-related libraries and approaches to create annotations. d3-annotation [42] is a library closer to visualization. It can onboard users to visualizations developed in D3.js [13]. Additionally, the libraries swoopyDrag [73] and labella [38] can create data-driven annotation and labeling for visualizations. These libraries are limited as they can create static labels in the visualization but cannot generate onboarding instructions, as we propose in this paper. Additionally, swoopyDrag [73] and labella [38] are no longer updated. These annotations libraries could be integrated into our proposed VisAhoi Library as an extension (see Figure 1). Besides annotations via programming, there are codeless tools such as chartAccent [54] and VisAnnotator [39]. ChartAccent [54] is a tool that allows users to quickly and easily augment charts via a palette of annotation interactions that generate manual- and data-driven annotations. Recently, Lai et al. [39] developed Vis-Annotator, an automatic annotating technique for promoting efficient data presentation with visualizations based on machine learning and a natural language interface. The presenter can upload a visualization image with the corresponding textual description and get a series of vivid well-annotated animations.

In summary, several products and services support developers in integrating onboarding elements. Approaches to automatically create annotations for visualization in the field of data stories are present in the literature [39, 32]. However, to our knowledge, none fully covers the onboarding requirements for data visualization, and a semi-automated creation of onboarding messages and in-place annotation anchors cannot be achieved.

## 3. Design Considerations

The presented onboarding concept in Section 2.4 is based on the results of previous studies [70, 65, 68, 67] and discussed in Section 2.3. In the following, we list the design considerations (DC) we derived and explain the design decisions when integrating visualization onboarding in a VA tool:

*Towards the Design of Visualization Onboarding*

- **DC1: Concrete examples.** Use concrete examples of how to read the chart [65, 70] by referring to the underlying data shown in the visualization. Our comparative user studies [70] found that concrete examples on how to read the chart are vital in increasing comprehension.
- **DC2: Visual Encoding & Interaction.** The generalization of the results of the related empirical studies and the design space [69] analysis shows that the visualization onboarding should explain the visual encoding & interaction concepts [2, 9, 37].
- **DC3: In-situ.** Use in-situ and to-the-point onboarding instructions [65, 70]. By integrating the onboarding system into the data visualization, users do not need to jump back and forth between onboarding and instructions. In addition, when using data visualization for insight generation, onboarding should be called up at any exploration stage. Qualitative feedback (see [70]) indicated that the to-the-point descriptions make it easier to absorb information. Likewise, the understanding of introductions is enabled by interactive and linked descriptions of the visualizations and the described steps.
- **DC4: Interactivity.** Provide interactively and linked onboarding instructions [70]. Users should be able to interact with the whole visualization interface when onboarding is activated.

Based on the design considerations, we derived objectives a visualization onboarding library needs to cover. We aim to integrate the visualization onboarding concept into VA tools using the underlying data of the visualizations. Available products for user onboarding mainly focus on the overall user interface rather than visual representations, as demonstrated by sources such as AppCues [4] and Intercom [33]. However, most user onboarding JavaScript libraries such as Hopscotch [31], aSimpleTour [5], intro.js [34], webTour.js [78], Tourguide.js [76], Orient [48], GuideChimp [27], tutoBox [77] are not fully suitable for visualization onboarding. This is because they are limited to guided tours that explain elements in the user interface, which can block the whole interface and prevent users from interacting with the visualization while seeking help. This is especially problematic when exploring visualizations because users should be able to interact with both the visualization and the onboarding simultaneously. Additionally, the existing onboarding libraries and solutions do not utilize the underlying data of the visualization, which is a critical factor in introducing unfamiliar visualizations to users. Hence, there is a need for a JavaScript library to integrate visualization onboarding into VA tools that utilize the visualization's underlying data to explain the visualization and not block the whole interface while using the onboarding instructions. Therefore, we implement a JavaScript library called VisAhoi, which is presented in this paper. We defined five objectives (OBJ) that a visualization onboarding library should meet from a developer's perspective.

*Towards the Design of a JavaScript Library for Visualization Onboarding*

- **OBJ1: Integrable.** Adding the onboarding library to existing visualizations and integrating it into an existing application plays an important role in gaining fast adoption and making visualization onboarding available for potential users of a VA tool. This also refers to DC3.
- **OBJ2: Extensible.** The onboarding library should be extensible to support multiple declarative visualization frameworks to provide broad coverage.
- **OBJ3: Reusable.** The onboarding library should provide reusable onboarding messages for different datasets for the same visualization type (e.g., horizon graph). Visualization designers/developers should be able to reuse onboarding messages across various visualization projects.
- **OBJ4: semi-based.** The template for the onboarding message should be provided with the onboarding library to generate onboarding instructions based on the given visualization type and dataset.
- **OBJ5: Customizable.** The developer should be able to customize the appearance of the onboarding messages as well as the annotation style.

This version of VisAhoi supports all five objectives. We discuss the limitations regarding these objectives in Section 6.2.

## 4. Visualization Onboarding Concept

The presented onboarding concept is based on the results of previous research by Stoiber et al. [70, 65] and has been actively developed further throughout the design process. The previously introduced onboarding concept, a step-by-step guide [70], could not be easily integrated into a VA tool because it was space-filling, with the onboarding instructions side-by-side with the visualization. Therefore, we surveyed different design alternatives on how to provide the onboarding messages more effectively. First, integrating them into the main visualization interface does not distract users from exploring the data. Second, this approach provides context-sensitive and to-the-point instructions. Therefore, in the current design, a floating action button [43] (Material Design Pattern) is used to access the onboarding instructions (see Figure 5 and Figure 7).

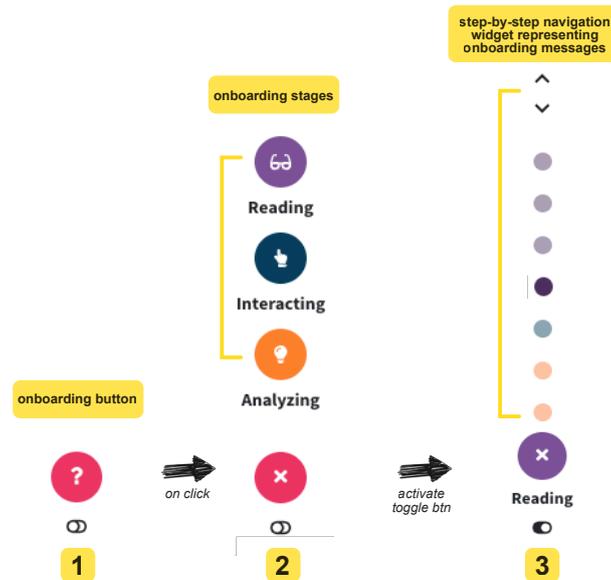

**Figure 5:** Onboarding button – Floating action button (1). The navigation comprises different elements to provide a guided tour and a self-exploratory experience. By clicking on the floating action button (1), the menu expands outward from a central point and reveals three onboarding stages: Reading, Interacting, and Analyzing (2). We integrated a step-by-step navigation widget that represents onboarding messages and arrow icons to navigate through them (3).

*Navigation:* In general, we designed the onboarding concept so that users can be either (1) guided through the different onboarding messages or (2) self-explore it and seek help at any stage of their interpretation and interaction with the visualization. At any point of exploration, the user can interact with the visualizations. Users are not restricted to step-wise navigation through the onboarding instructions as known from other user onboarding libraries [31, 5, 34, 78, 76, 48, 27]. The onboarding concept provides context-sensitive help, a main design goal when integrating onboarding into VA tools (cf. Stoiber [69]). The main navigation element is a floating action button (Material Design Pattern) [43]. By default, we positioned it on the lower right side of the visualization, as illustrated in Figure 7. Thus, we can independently add onboarding instructions to different visualizations. Clicking on the button reveals three circular buttons, representing the three default stages of the onboarding process: Reading, Interacting, and Analyzing (Figure 5). In the onboarding stage, *Reading the Chart* onboarding instructions explain the visual encoding (e.g., size, color); *Interacting with the Chart* onboards the user to the interaction concept. Moreover, *Analyzing the Chart* guides the user towards further insights (e.g., making comparisons). The step-by-step navigation widget represents the number of onboarding messages to implement the guided approach (Figure 5 (3)). Above the onboarding button, illustrated in Figure 5 (3), up and down arrow icons allow the user to browse through the onboarding messages. Users who seek help for a specific part of the visualization, e.g., interaction, can self-explore the onboarding messages by opening a stage (Reading, Interacting, or Analyzing) in the main navigation menu and then clicking on the specific in-place annotation anchor in the visualization.

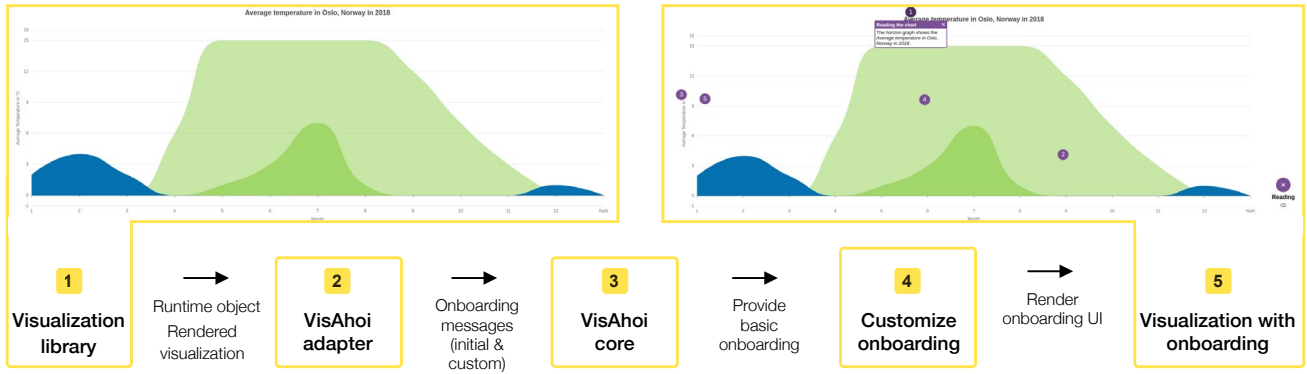

**Figure 6:** The VisAhoi workflow: (1) a visualization generated by a visualization library; the library renders a visualization from a given dataset, visualization specification, and DOM node and provides a runtime object, (2) the VisAhoi adapter specific to the visualization library extracts information, e.g., values and anchors, from the visualization specification and runtime object, and provides initial onboarding messages that can optionally be customized, (3) the VisAhoi core generates onboarding messages; The values are applied to the onboarding messages. Due to missing values or anchors, incomplete messages are filtered out, (4) the onboarding can be further customized in terms of visual appearance, and (5) VisAhoi renders the onboarding interface with in-place annotation anchors at the given position in the visualization.

*Onboarding messages:* Assumptions about what type of example datasets are best suited to use for onboarding when explaining visualization systems and the underlying data characteristics tend to vary widely. On the one hand, literature calls for abstract data to be more generalizable and transferable to a novel context [35]. On the other hand, more recent scientific work provides counter-evidence. For instance, De Bock et al. [19] show a more successful knowledge transfer using concrete examples. Similarly, when introducing new visualizations, it is recommended to use an accessible and understandable dataset that can be assumed to be well-known by the general public [40, 36, 26]. A comparative study by Stoiber et al. [65] between abstract and concrete onboarding messages for a treemap visualization showed that both concrete and abstract onboarding messages could lead to valuable insights. The VisAhoi library provides abstract onboarding messages by default, which are semi-automatically generated based on the underlying data of the visualization. For the horizon graph, for example, default messages in the reading stage look like *The horizon graph shows the Average temperature in Oslo, Norway in 2018.* and *Light green areas indicate a moderate positive Average temperature in °C and dark green areas a high positive Average temperature in °C.*. Access the demo page for more examples here: https://datavisyn.github.io/visAhoi/demos/. For our usage scenarios, we provide concrete onboarding messages for biomedical researchers, as described in Section 6.1. We created the onboarding messages in close collaboration with a domain expert. For the second usage scenario in data-driven journalism, we provide abstract onboarding messages for an authoring tool designed for data journalists (see Section 6.2). Onboarding messages can be relocated from their initial position by the user with drag & drop in order not to cover essential visualization elements.

*In-place annotations for anchoring onboarding messages:* Textual instructions (onboarding messages) are shown in the form of tooltips in combination with in-place annotation anchors [84] (numbers and symbols) to indicate the connection between the text elements and the visual encoding as shown in Figure 7 (3). Inspired by different annotation designs by Lu [42], we use circular anchors with numbers related to the onboarding message.

In the following section, we describe the design decisions of the VisAhoi development, motivated by our objectives (see Section 2.4).

## 5. Design of the VisAhoi Library

The main features of the VisAhoi library are (1) the generation of visualization onboarding instructions using a web-based declarative visualization library, which is semi-automatically generated based on the specification; (2) the customization possibilities of all the elements of the visualization onboarding concept, which we describe in Section 5.4. In particular, VisAhoi supports web-based declarative visualization libraries, which create the visualization based on a specification (e.g., JSON) that describes the visual encoding, the interactive behavior, and the underlying

```
1  import { ahoi, EVisualizationType } from '@visahoi/plotly';
2  // create plotly visualization
3  const runtimeObject = await new Plotly.newPlot(element, traces, layout);
4  // initialize visAhoi
5  const visAhoi = ahoi({
6    visType: EVisualizationType.SCATTERPLOT,
7    chart: runtimeObject,
8    ahoiConfig: {}, // additional configuration
9    customizeOnboardingMessages: (defaultOnboardingMessages: IOnboardingMessage[], contextKey: string) =>
10     customizeHexbinplotMessages(defaultOnboardingMessages, contextKey, runtimeObject, numberOfDatapoints),
11 });
```

Listing 1: Each visualization library adapter exports an `ahoi()` function that takes care of extracting the information from the runtime object for the supported visualization types and generates the onboarding. This listing shows an example for initializing `visAhoi` for a Plotly.js visualization.

data of the visualization. Those libraries transform the input into a library-specific runtime object and render the visualization in the provided DOM node.

VisAhoi is developed using the JavaScript framework Svelte [71]. We chose Svelte, as it compiles plain JavaScript and allows developers to integrate VisAhoi in applications with different frontend frameworks. VisAhoi currently supports three popular visualization libraries: Plotly.js [50], Vega-Lite [57], and Apache ECharts [41]. The VisAhoi repository contains the VisAhoi adaptors for each supported visualization library and the VisAhoi core, which contains functionality that can be shared independently of the visualization library. An additional demo package is included to demonstrate the VisAhoi integration for different visualization types for each supported visualization library (https://datavisyn.github.io/visAhoi/demos/). The current implementation of our VisAhoi library supports the following visualization types: scatterplot, bar chart, change matrix/heatmap, treemap, and horizon graph. A generic type is available so that developers are not limited to the supported visualization techniques of the library. The generic type initializes the onboarding with an empty set of onboarding messages, which the developers can then define.

To initialize VisAhoi, developers only need to fetch the runtime object of the created visualization and pass it on to VisAhoi when calling the `ahoi()` function, together with the visualization type. The `ahoi()` function provides a callback function (`customizeOnboardingMessages()`) as a parameter, which provides the default onboarding messages created by VisAhoi for the specific visualization type. Developers can either rely on those, customize them, or add new onboarding messages (see Section 5.4). Every onboarding message needs to be assigned to an onboarding stage, which is basically the group where all related onboarding messages will be rendered. All onboarding messages will then be added to the state of VisAhoi and, finally, get rendered in the specified onboarding stage in the onboarding user interface (UI) created by VisAhoi.

In the following, we explain the VisAhoi architecture in more detail.

### 5.1. VisAhoi Architecture

The library consists of two main components: the visualization-library-specific adapter and the core, illustrated in Figure 6 (2) and (3).

*VisAhoi adapter:* The VisAhoi adapter is visualization-library specific, which means we provide one for Plotly.js, Apache ECharts, and Vega-Lite, each. It considers the DOM architecture of the visualization and the specification object, which differ for each visualization-specific library. Depending on which library developers use, they need to import the respective package. When developers initialize the onboarding by calling `ahoi ()`, showing in Listing 5.1, the function in the respective package is called, to which the visualization type and the runtime object need to be passed. Additionally, a configuration object and a function to customize the onboarding messages can be passed if the developer needs further configuration or customization of the onboarding (see Section 5.4).

The adapter provides functionality for each supported visualization type. It creates the default onboarding messages by considering the visualization type and extracting the necessary information from the runtime object. The runtime object contains the DOM element, the chart specification, and the data contained in the visualization, which is used to get the necessary information for individual in-place annotations.

Each visualization type-specific function across all visualization library-specific adapters returns an object, the onboarding specification, with the same structure, which will later be passed on to our core package. The onboarding specification contains information about how to extract specific information from various visualization elements and

```
1   chart title: {
2       value: runtimeObject?.layout?.title?.text,
3       anchor: {
4         findDomNodeByValue: true,
5         offset: { left: -20, top: 10 },
6       },
7   },
8
9   yMin: {
10      value: min.toFixed(2),
11      anchor: {
12        coords: {
13          x: xGrids[minIndex]?.getBoundingClientRect().x,
14          y: traceNodes[2].childNodes[0].getBoundingClientRect().y,
15        },
16      },
17  },
18  yAxis: {
19      value: chart.layout.yaxis.title.text,
20      anchor: {
21        sel: ".infolayer .ytitle",
22      },
23  },
```

Listing 2: This listing shows the positioning of the in-place annotation anchors exemplary for the chart title and y-axis.

where to place the respective anchor. Listing 5.1 shows an example of extracting information for the visualization title and the minimum value along the y-axis.

In-place annotation anchors can be positioned in two ways. The information is stored in the anchor attribute, which you can see in Listing 5.1 in lines 3, 11, and 21. On the one hand, the anchor will be absolutely positioned in the visualization (see Listing 5.1, yMin). On the other hand, the anchor can be attached to a DOM node inside the visualization. To attach the anchor to a DOM node, we again have implemented two different approaches. Either VisAhoi looks for the DOM node by the given content, e.g., the text in the visualization title (see Listing 5.1, chartTitle). To do that, the flag `findDomNodeByValue` in the anchor object needs to be set. Or, the selector is defined, which VisAhoi will then be used to find the corresponding DOM node in the visualization (see Listing 5.1, yAxis, line 21). We generally recommend the latter approach, as we made the experience that DOM elements inside the visualization can have the same text content in some cases, which would lead to wrongly positioned anchors. The created onboarding specification for the current visualization is then passed on to the core package, where the onboarding messages will be created.

*Core:* Our core package includes functionality that is shared across different visualization libraries. When the onboarding specification is created in the corresponding adapter, it will be passed to the core, where the onboarding messages will be created (see Figure 6). Onboarding messages will be the same across different visualization libraries. In case the adapter cannot correctly extract all the necessary information from the runtime object, the onboarding messages cannot be displayed. The core method `generateMessages()` receives the runtime object, the onboarding specification, and the visualization type. Depending on the visualization type, onboarding messages will be created appropriately. To display a specific message, specific information about the visualization is needed. Depending on whether VisAhoi has this information or not, the onboarding message will be rendered in the end. For example, to explain the visualization title, finding the visualization title inside the visualization is necessary. If the adapter is not able to extract this information from the runtime object, the anchor and onboarding message for the visualization title will not be rendered. The core package defines which information is needed for each attribute of the visualization. Listing 5.1 shows an example of this object for the visualization title. The attribute `requires` in line 3 defines which properties are needed from the onboarding specification, which is passed from the adapter. In this case, the attribute `charttitle`, defined in the adapter (see listings 5.1 and 5.1), is needed to display the onboarding message in the UI.

Eventually, the `generateMessages()` function returns an object containing all possible onboarding messages for the specific visualization type. Those messages will be passed back to the developer through the callback function `customizeOnboardingMessages()` mentioned above, where they can be filtered, updated, extended, and customized.

```
1  {
2  anchor: getAnchor(spec.chartTitle, visElement),
3      requires: ['chartTitle'],
4      text: `The horizon graph shows the <i>${spec.chartTitle?.value}</i>.`,
5      title: 'Reading the chart',
6      onboardingStage: reading,
7      marker: {
8        id: `visahoi-marker-${contextKey}-8`
9      },
10     id: `visahoi-message-${contextKey}-8`,
11     order: 1
12 }
```

Listing 3: The core package of VisAhoi defines requirements for the default onboarding messages. In this case, to show the onboarding message for the visualization title, the attribute `chartTitle` needs to be given in the onboarding specification that the core gets passed from the adapter (see listing 5.1).

### 5.2. Internal State of VisAhoi

As VisAhoi is developed using the frontend framework Svelte, it also uses a Svelte state internally to store all the information for onboarding. It is necessary to have a state per visualization because everything related to the onboarding is stored in this state, starting from the chart-specific onboarding messages or stages and any custom configuration. We solved this by having a global state that holds all states of all visualizations as a key-value pair. The key (called `contextKey`) is the unique identifier that is used to find the associated onboarding for a visualization. Developers can either pass their own `contextKey` when initializing the onboarding by passing it to the `ahoi()` function. In case no `contextKey` is passed, VisAhoi will create one and return it back to the developer. Whenever customizing or retrieving the onboarding of a visualization, this key needs to be passed to the respective VisAhoi function. Having this key, VisAhoi will then store the update to the corresponding internal state of the visualization.

### 5.3. User interface of VisAhoi

Onboarding stages, onboarding messages, the in-place annotation anchor, the tooltip, and the onboarding navigation are part of the core package of VisAhoi and can be customized by the developer. When the adapter extracts all necessary information, it passes the onboarding specification onto the core package, generating the onboarding messages. Those messages are the basis and will be used to render the onboarding in the DOM finally. Therefore, VisAhoi creates an SVG element with the same width and height as the visualization. This information can be extracted from the passed runtime object. The created SVG element will be positioned exactly on top of the visualization so that the onboarding, which will be rendered inside, exactly overlaps the visualization. This way, the following described UI elements of the VisAhoi interface are rendered correctly. To keep the possibility to interact with the original visualization as the SVG of the onboarding is rendered right above, we added `pointer-events: none`. This way, users can interact with the visualization even when the onboarding is opened.

### 5.4. Customization

VisAhoi allows a variety of customization options for the developer. It allows adapting the general UI of the onboarding, including icons and colors, to make it easy to integrate onboarding into a given visual analytics tool.
**Onboarding stage:** By default, VisAhoi provides three onboarding stages: Reading, Analyzing, and Interacting with the chart. The developers can adjust onboarding stages by changing the color, the icon, and the stage title or even deleting or adding new onboarding stages.
**Onboarding message:** The onboarding message consists of a title, text, an anchor (=position where it should be placed in the visualization), an in-place annotation (=annotation in the chart), an order (order within an onboarding stage), a tooltip position (should the tooltip be rendered on top/bottom/left/right of the anchor), and required information to display the message. Additionally, an onboarding message needs to have a corresponding onboarding stage. The developer can customize onboarding messages: Changing the onboarding instructions (text), order, and corresponding onboarding stage.
**Onboarding navigation:** Developers can decide if users should be able to show the stepper navigation in the onboarding navigation. The position of the onboarding navigation can also be customized: it can be unfolded upwards or to the left.

**In-place annotation anchor:** In-place annotations are attached to visualization elements to highlight that onboarding is available. In-place annotation anchors can have text, e.g., numbers, guiding the user through onboarding. The developer can decide if the numbers are needed.
**Tooltip:** Tooltips are attached to in-place annotation anchors and reveal the onboarding information when clicking on the anchor. The tooltip includes a title and the onboarding message, which the developer can adjust. Furthermore, the developer can highlight essential words in the onboarding message by formatting it using HTML.

All these configuration options can be passed to the `ahoiConfig` object in the `ahoi()` function when initializing the onboarding for a visualization.

### 5.5. Inline-edit mode

VisAhoi provides an edit mode for users of visualizations where onboarding is available. This feature was specifically implemented for the data-journalistic use case with Flourish (see Section 6.2). VisAhoi provides a function `setEditMode()` that enables developers to enable or disable the edit mode. When enabled, the user interface of the onboarding of VisAhoi changes slightly. Users, e.g., of an authoring tool, can adjust the onboarding, for example, edit or delete the onboarding message, edit the tooltip title, or delete an onboarding stage inline. It is not recommended to generally enable this feature for all use cases as users could edit the onboarding of visualizations they are not familiar with.

*Summary:* VisAhoi generates onboarding messages semi-automatically by extracting information, e.g., values, anchors from the visualization specification and runtime object, and provides default onboarding messages that the developer can customize. The onboarding can be further customized by changing onboarding messages, onboarding stages, titles, icons, and colors of the onboarding concept. The library allows developers to integrate onboarding instructions into VA tools, extend it as it uses declarative visualization frameworks, reuse onboarding messages across visualization projects, semi-automate the generation of onboarding messages, and highly customize it. In the next section, we describe in two usage scenarios how to integrate the library into a VA tool for the analysis of high-throughput screening (HTS) data for biomedical R&D, and into a Flourish template to provide a visual authoring tool for data journalists to provide onboarding for a treemap visualization.

## 6. Usage Scenario

We present two usage scenarios [59] showing the applicability of the VisAhoi library in biomedical R&D and data-driven journalism.

### 6.1. Integration into a VA tool for HTS data analysis

The VA tool that VisAhoi is integrated into is used to analyze biomedical high-throughput screening (HTS) data, illustrated in Figure 7. It holds data from HTS experiments where the binding kinetics of roughly four million chemical compounds are measured against potential drug targets. The measurement for each compound is a time series curve that describes the binding behavior of the compound against the target over time. The four million time series measurements are projected into a two-dimensional latent space with a multidimensional scaling approach. The 2D projection is visualized with a hexbin plot, which bins single points close to each other to girded hexagons to prevent overplotting. The density of single points behind a hexagon is encoded in color. Individual hexagons can be dissolved to show the actual compounds as scatter points behind them. The topology in the 2D space reflects the topology of the time series curves in the high-dimensional space. The tool offers a scatter plot on the right side, where each dot represents a single compound. The attributes shown on the x and y axis and the color encoding are fully customizable by the user. The plots are implemented as multiple-coordinated views, which means a selection in one of the plots is highlighted in the other. Domain experts use these plots to search for clusters of compounds showing promising binding behavior. Both plots are built with the plotting library Plotly.js.

Domain experts that use this VA tool used to work with the scatterplot on the right side for many years. The latent space embedding (left) is a new technique for most users. Therefore, onboarding is not needed to explain what a scatterplot is, but rather what the data means: How was it pre-processed, and how to draw conclusions properly. Furthermore, domain experts are used to reading from static plots, whereas this VA tool offers many interaction possibilities to dive deep into the visualizations. To get the best out of this tool, additional onboarding is needed to explain how to interact with and analyze the given visualization, at least for the first few times the user works with this tool.

### I  *VisAhoi* integrated into VA tool for HTS data analysis

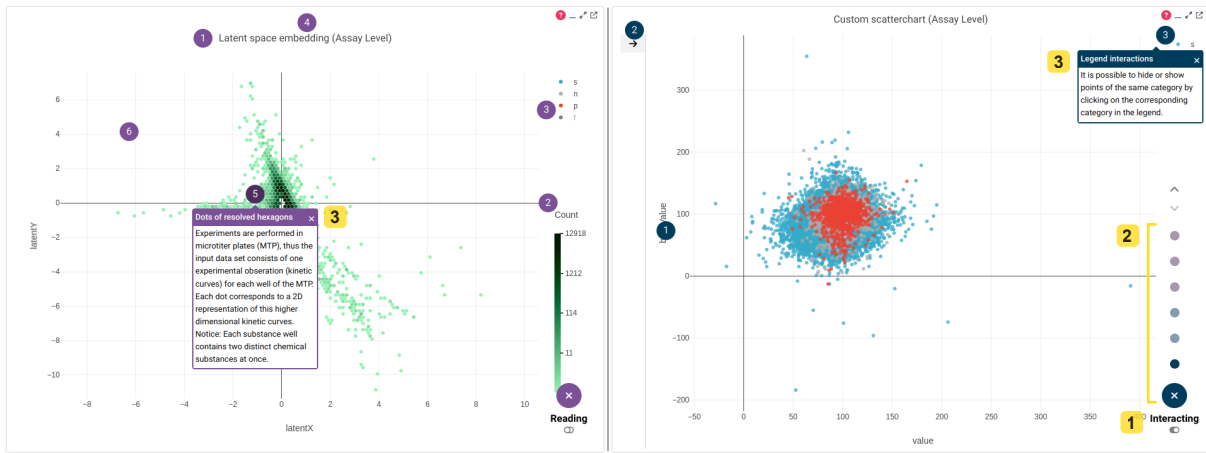

### II  Authoring Tool *VisAhoi* integrated into Flourish

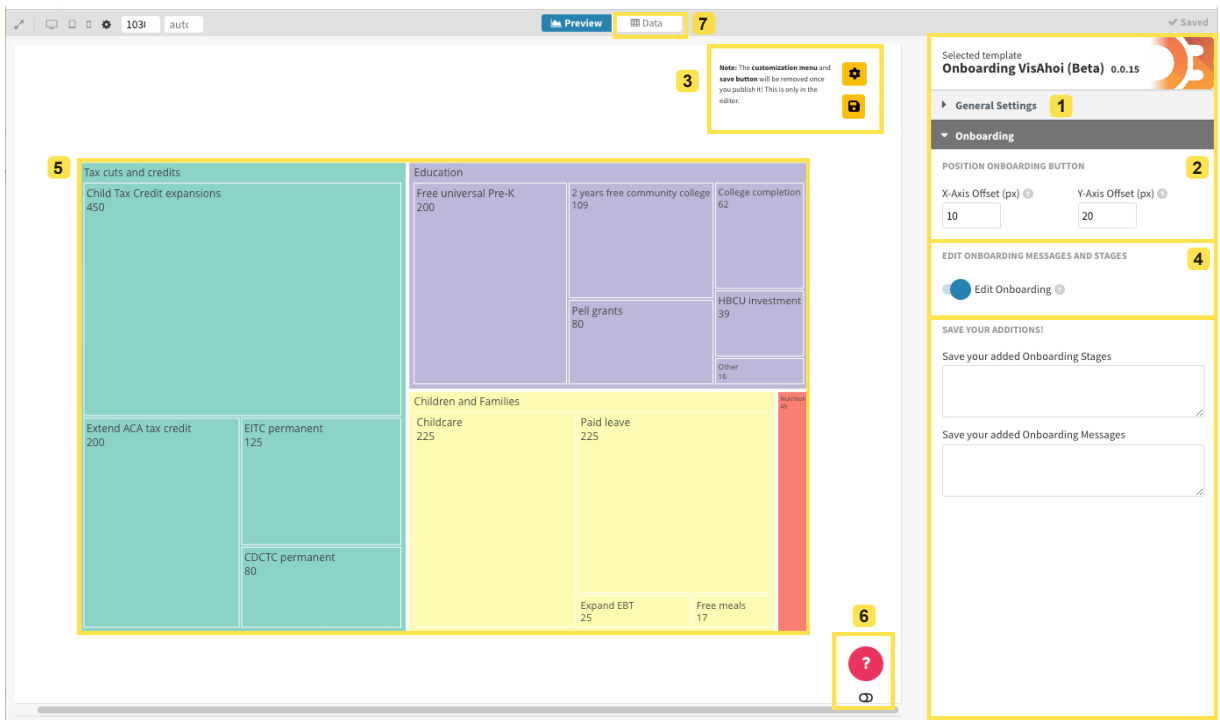

**Figure 7:** Usage scenarios for VisAhoi integrations. **(I) VisAhoi integrated into the VA tool for analyzing HTS data** — side-by-side a hexbin plot (left) and scatterplot (right). The navigation comprises different elements to provide a guided tour and a self-exploratory experience (II 1, 2). By clicking on the floating action button (1), the menu expands outward from a central point and reveals three onboarding stages: Reading, Interacting, and Analyzing. We decided to integrate a step-by-step navigation widget (2) that represents onboarding messages (3) and arrow icons to navigate through them (2). The onboarding messages are designed as tooltips with in-place annotations (4) to indicate the connection between the textual instructions and the visual encoding. The user can open both onboarding instructions for both visualizations simultaneously. **(II) VisAhoi integrated into a treemap authoring tool for data journalists.** — the interface is divided into two areas: (A) preview with the visualization (plotly.js) (5), the onboarding button (provided by the VisAhoi library (6), and the settings panel, and (B) Flourish (1) with the settings regarding onboarding customizations (2, 3, 4).

*Integration of VisAhoi:* The developer integrating VisAhoi in the VA tool was an experienced full-stack developer who gained experience in various languages over the past years. She is very experienced in the use case relevant programming language, Typescript.

For this particular usage scenario, it is important that onboarding messages can be customized, and the developers themselves can add additional ones as domain experts already know how to read a scatterplot in general. As described above, we focused on explaining how the data was pre-processed, explaining the hexbin plot, and how the multi-coordinated view and its interaction work. So we started by taking a look at the semi-automatically generated onboarding messages by the library that are provided as a parameter of the `ahoi()` function, called `customizeOnboardingMessages()`. We decided that only the onboarding messages of the stage *Interacting with the chart*, which describe the chart interactions, are relevant for this use case. Hence, we removed all other messages. We also customized messages for the *Interacting with the chart* stage to fit our usage scenario perfectly. Furthermore, we created messages for the two stages *Reading the chart* and *Analyzing the chart* by calling the method `createBasicOnboardingMessage()` for each onboarding message. We pushed all messages to an array that now holds the basic, customized, and newly added messages. All those messages are finally returned from the callback function `customizeOnboardingMessages()` in the `ahoi()` function, which will be passed to VisAhoi later on.

We packed all this functionality into a function in our code called `customizeHexbinplotMessages()` and `customizeScatterplotMessages()`, which can be re-used anytime.

To initialize the onboarding for our two visualizations, we call the `ahoi()` function, which is exposed by the VisAhoi library, and pass the visualization type (scatterplot in both cases), the runtime object, and the function that returns the customized onboarding messages (see listing Listing 5.1). `ahoi()` returns the onboarding instance, enabling us to show and hide the onboarding as the user interacts with the application. We added a button to our user interface that hides or shows the VisAhoi onboarding on click (see pink "?" icon in the top right corner of the visualizations in Figure 7). This way, new users of the tool can easily access the onboarding on demand, and power users do not feel disturbed by it.

*Lessons Learned:* In summary, integrating VisAhoi into the VA tool was straightforward. A few lines of code can add the basic onboarding to visualizations. Having basic onboarding messages was helpful as it allowed the developer to see what the onboarding could look like. This is a good starting point and makes it easier to decide what to customize in the next step. As the default onboarding messages were not enough for the use case, additional ones have been added. This was not an issue as they can be passed to the `ahoi()` function as an additional parameter. To find the right position for the anchor of new messages, the developer has to dive deeper into the runtime object of the visualization. As the visualizations of this VA tool hold huge amounts of data, canvas support (discussed in section 7.3) would be helpful to annotate individual data points better (e.g., min/max values along an axis), which is currently only supported for SVG. Nevertheless, the developer placed all needed onboarding messages properly in the visualization.

### 6.2. Combining VisAhoi & Flourish in an authoring tool for (data) journalists

The second usage scenario was performed in data-driven journalism [68]. Many journalists lack programming skills [53, p.8], they use tools such as Flourish [23], Datawrapper [18], and Tableau [74] to create their data visualizations for data stories [1]. *Flourish* [1] only partially integrates some of the above-mentioned onboarding approaches, such as annotations or storytelling. These features must be integrated manually by the user or restricted to a specific type of data visualization (e.g., line charts, bar charts, or pie charts). The visualization software provides stories [25] that let the journalists create richer and more in-depth data narratives with animated transitions. Additionally, annotations [24] can be added as labels, blocks of text, or connector lines to particular sets of data visualizations. However, these approaches exhibit weaknesses as they have to be designed fully manually. They are unrelated to the data in the data visualization, cannot be used as a guided tour, and affect more experienced users in their exploration as they are always visible. Therefore, we aimed to enhance traditional Flourish templates with an authoring tool (targeted towards (data) journalists) by combining our self-developed VisAhoi library and the Flourish template to support data journalists semi-automatically generating visualization onboarding concepts that support readers in interpreting more complex visualization techniques.

*Integration of VisAhoi in Flourish:* The authoring tool provides a user interface to add onboarding and customize it: (1) editing, reordering, and/or removing the semi-automatically generated onboarding messages and onboarding stages, and (2) adding own messages and stages. The prototype can be accessed at http://bit.ly/3gBqrxv. The journalist can activate an edit mode in the Flourish template and adjust the onboarding visually using the authoring user interface. We used a treemap with Biden's tax overhaul [52] data as an example data set in our usage scenario implementation. Treemaps [63] are well suited to represent hierarchical data with a quantitative attribute. However, treemaps are not particularly well-known among audiences [12] and might be hard to interpret initially. Therefore, we decided to show the functionality of our authoring tool for a treemap visualization.

Over the last two and a half years, we collaborated closely with Austrian data journalists, thereby iteratively developing a better understanding of the problem and designing and evaluating the authoring tool to address it.

The developer of the authoring tool closely worked together with the colleagues developing the VisAhoi library. He has experience in front-and backend development with various languages, mostly JavaScript, for over ten years. Working with the VisAhoi library and personal experience will be further elaborated in *Lessons Learned*.

Plotly.js [50] is used to render the treemap visualization, passing the runtime object to the VisAhoi library to create the semi-automatically generated onboarding instructions. Several internal functions of VisAhoi were used to realize this authoring tool, and also the `setEditMode()`, which enables the visual configuration of the onboarding within the Flourish template by the user.

The onboarding messages and stages are generated semi-automatically by default based on the visualization and the data. We provide a more generic description of the treemap, which does not include any concrete values from the data visualization, e.g., *"The size of each rectangle represents a quantitative value associated with each element in the hierarchy."*. Based on our comparative study [65] between abstract and concrete onboarding messages for a treemap visualization, we found that both concrete and abstract onboarding messages can lead to highly valuable insights. Therefore, we included abstract onboarding messages to provide a blueprint for the journalists, which can be reused and personalized for other treemap visualizations.

The authoring tool uses mostly Bootstrap [11] for the styling and the modal windows. While implementing the Flourish template, we had some technical restrictions. The settings panel can only be customized with predefined elements through a *.yml* file. To include our custom overlay menu, we had to place it in the visualization area. However, the menu is only visible in the "editor view" of the template. We used the settings menu behavior to automatically save the customizations of the onboarding once the user interacts with one of the controls in the settings panel of Flourish. We provided two text fields — for the onboarding stages and messages separately — where the user had to copy a JSON string automatically generated when pressing the "save" button. The separation of the two is necessary to restore the previous state of the Flourish template when reloaded.

*Lessons Learned:* The utilization of the VisAhoi library and its integration with Plotly was straightforward and efficient. The generation of the visualization with the automated onboarding worked without any issues. However, adding custom messages was more challenging due to Flourish's SDK. Despite the difficulties, the library's customization functions were user-friendly once operational. One significant advantage of the library was that state saving within the app was not necessary since the library automatically saved the onboarding messages' state.

## 7. Discussion, Lessons Learned, & Future Work

In this section, we discuss the result of our two usage scenarios, present lessons learned, and present limitations of the VisAhoi library. Furthermore, we also provide future directions.

### 7.1. Lessons Learned

*VisAhoi is easily integrateable.* VisAhoi is a Typescript library designed with a developer's experience in mind. Adding semi-automatically generated onboarding to visualizations can be done with a few lines of code (see Listing 5.1) (Objective 2.4: Integratable). The basic onboarding messages, semi-automatically generated by VisAhoi, can be used as a blueprint to extend and further adjust towards users' needs (see Section 2.4).

*Customization possibility for the developer.* The customization possibility (Objective 2.4: Customizable) of VisAhoi allows developers and visualization designers to flexibly adjust the onboarding concept to meet the user's needs of the VA tool. The usage scenario showed that customization possibilities, e.g., adjusting onboarding messages and stages, are essential for domain experts' needs to adjust the onboarding concept. Biomedical scientists are used to interpreting scatterplot visualizations illustrated in Figure 7 (I.). Therefore, the developer only provides onboarding instructions for *Reading the chart* and *Interacting with the chart*. As the concept of a hexbin plot and the contained data is new for the users of this application, we provide onboarding messages for all three onboarding stages — *Reading, Interacting, and Analyzing* — to support the user. Furthermore, the inline-edit mode, which the VisAhoi library provides, allowed us to develop the authoring tool for data journalists (see Section 6.2). The provided function `setEditMode()` enables users to edit or delete the onboarding message, the tooltip title, an onboarding stage, etc., via the UI.

*VisAhoi supports reusability of onboarding messages.* VisAhoi is reusable as developers can reuse onboarding messages. Developers can access the state of the onboarding for all visualizations using the context key (Objective 2.4: Reusable). They can also retrieve the onboarding messages from VisAhoi and store them in their application state to reuse them for other visualizations in the same application.

*Extensibility is supported by a generic type.* VisAhoi is extensible (Objective 2.4: Extensible). First, it supports three different visualization libraries Vega-Lite, Plotly.js, and Apache ECharts. Second, different visualization techniques are supported, such as bar charts, change matrices, horizon graphs, scatterplots, and treemaps. We showed the extensibility of the VisAhoi library in the usage scenarios in Section 6. We also implemented a generic type to provide onboarding for other visualization techniques. The only restriction is that there will not be any semi-automatically generated onboarding messages available to start with. The developer has to define them manually.

*The specification of the declarative visualization library (grammar) is insufficient to provide valuable visualization onboarding messages.* While developing our VisAhoi JavaScript library, we realized that declarative grammar alone could not provide valuable onboarding messages. For example, information like the color of a single rectangle in a heatmap cannot be retrieved from the grammar. The grammar only holds the raw data values and the color for minimum and maximum values, but the mapping of the data values and the corresponding color can only be retrieved via the runtime object. Another example of why grammar alone is insufficient is the position where to place the marker. Some markers are attached to specific DOM nodes of the visualization, e.g., the title, which is a piece of information we cannot receive from the grammar alone but need the rendered visualization or the runtime object for. Therefore, we use the runtime object of the visualization library to extract all vital information to generate onboarding messages semi-automatically.

### 7.2. Limitation

A general limitation of VisAhoi is the fragility of the visualization adapters, as they depend on the runtime objects of the visualization libraries. The adapter might break or return different onboarding messages with different versions of the visualization library when the internal data structure changes. As mentioned in Section 5, we require the runtime object to access the actual values used to render the visualization. A solution for this limitation would be a standardized application programming interface (API) for runtime objects that the visualization libraries provide.

### 7.3. Future Work

*Canvas Support.* VisAhoi explicitly targets visualization libraries with SVG rendering, which facilitates the selection of anchors for the in-place annotations by employing the DOM structure. In contrast, canvas rendering has a better performance but does not provide additional structure. The anchor coordinates must be retrieved exclusively from the runtime objects that often lack this information. Thus, we positioned the onboarding message manually in the hexbin plot to provide the instruction for the position and meaning of the hexagon, as we could not retrieve the DOM element due to the canvas restriction. In a future iteration, we want to analyze the canvas renderer of the three selected visualization libraries. The approach could then be applied to other libraries, such as the popular Chart.js [17] library, which uses canvas rendering only.

*Onboarding for Interactions.* Explaining available user interactions with a visualization, such as, for instance, brushing or item selection, is a major aspect besides visual encoding. In contrast to visual encoding, the declaration of interactions differs between visualization libraries and can be more complex in multiple coordinated views (MCV). ECharts and Vega-Lite describe interactions as part of the visualization specification [57]. The visualization library adapter could identify and extract the information necessary to generate matching onboarding messages. For Plotly.js, VisAhoi provides onboarding messages for all interaction possibilities in the modebar. When creating a single visualization, developers can decide which interactions to allow and show in the modebar and VisAhoi adapts the corresponding onboarding message. In a MCV setup, Plotly.js requires custom JavaScript code using Plotly's selection event listener for MCV with linked visualizations. Hence, it is more challenging to implement onboarding messages for custom event listeners than to build a functionality that can be accessed using the runtime objects.

*Data Insights.* With the visualization onboarding, the infrastructure can provide insights into the dataset using quality metrics [8, 7]. Quality metrics can reveal one or multiple interpretable visual patterns for a specific visualization type and dataset. Onboarding can explain the findings to the user in a comprehensive manner. Our prototype provides a glimpse of this feature by displaying extracted statistical values from the visualization library, e.g., minimum and maximum values. Showing onboarding messages for data insights could be especially useful in combination with the user interaction, creating a feedback loop that could guide users in their visual analysis process [15]. Once the user interacts with the visualization, the quality metric is re-computed. It reveals new, potentially exciting patterns that can be presented to the user as onboarding messages and/or highlights. This aligns with our vision, where users get an onboarding tailored to a visualization and a dataset rather than a generic tutorial.

## 8. Conclusion

We presented VisAhoi, a JavaScript library that semi-automatically extracts and describes onboarding annotations in the respective visualization using the high-level declarative visualization grammars Vega-Lite, Plotly.js, and ECharts. The VisAhoi library is easy to integrate and extensible, as the developers can write their own adapters to use further visualization libraries. Besides, the library is reusable and highly customizable. The current prototype demonstrates that it can generate and integrate onboarding instructions semi-automatically, which provides a solid foundation for future developments.

## Acknowledgments

This work was funded by the BMK under the ICT of the Future program via the SEVA project (no. 874018) and by the Austrian Science Fund as part of the Vis4Schools project (I 5622-N) and the docs.funds.connect project Human-Centered Artificial Intelligence (no. DFH 23-N). For the purpose of open access, the author has applied a CC BY public copyright licence to any Author Accepted Manuscript version arising from this submission. This work was additionally supported in part by the FFG, Contract No. 881844: "Pro²Future is funded within the Austrian COMET Program Competence Centers for Excellent Technologies under the auspices of the Austrian Federal Ministry for Climate Action, Environment, Energy, Mobility, Innovation and Technology, the Austrian Federal Ministry for Digital and Economic Affairs and of the Provinces of Upper Austria and Styria. COMET is managed by the Austrian Research Promotion Agency FFG.

VisAhoi: A Library to Generate Visualization Onboarding